\documentclass[conference]{IEEEtran}

\usepackage{graphicx,amsmath,amssymb}
\usepackage{subfigure}
\usepackage{cite}
\usepackage{epstopdf}
\usepackage{fancyhdr}
\usepackage{mdwmath}
\usepackage{mdwtab}
\usepackage{balance}
\usepackage{xcolor}
\usepackage{bm}
\usepackage{amsthm}
\usepackage{algorithm}
\usepackage{multirow}
\usepackage{flafter}
\usepackage{setspace}
\usepackage{booktabs}
\usepackage{tabularx}
\usepackage{threeparttable}
\usepackage{stfloats}
\usepackage{algorithmic}
\usepackage{mathrsfs}
\usepackage{enumerate}
\usepackage{url}
\usepackage{caption}
\usepackage{lipsum}
\usepackage{gensymb}

\usepackage[
top    = 1.81cm,
bottom = 1.05in,
left   = 0.6 in,
right  = 0.6 in]{geometry}

\allowdisplaybreaks[4]

\newtheorem{lemma}{Lemma}

\newtheorem{corollary}{Corollary}

\newtheorem{assumption}{Assumption}

\newtheorem{guideline}{Guideline}

\newtheorem{definition}{Definition}

\begin{document}
\begin{sloppypar}

\title{Coverage and Capacity Optimization in STAR-RISs Assisted Networks: A Machine Learning Approach}

\author{\IEEEauthorblockN{Xinyu~Gao\IEEEauthorrefmark{1}, Wenqiang~Yi\IEEEauthorrefmark{1}, Alexandros~Agapitos\IEEEauthorrefmark{2}, Hao~Wang\IEEEauthorrefmark{2}, and Yuanwei~Liu\IEEEauthorrefmark{1}, }

\IEEEauthorblockA{\IEEEauthorrefmark{1} Queen Mary University of London, London, UK\\
\IEEEauthorrefmark{2} Ireland Research Center, Huawei Technologies, Dublin, Ireland\\
}}

\maketitle

\begin{abstract}
  Coverage and capacity are the important metrics for performance evaluation in wireless networks, while the coverage and capacity have several conflicting relationships, e.g. high transmit power contributes to large coverage but high inter-cell interference reduces the capacity performance. Therefore, in order to strike a balance between the coverage and capacity, a novel model is proposed for the coverage and capacity optimization of simultaneously transmitting and reflecting reconfigurable intelligent surfaces (STAR-RISs) assisted networks. To solve the coverage and capacity optimization (CCO) problem, a machine learning-based multi-objective optimization algorithm, i.e., the multi-objective proximal policy optimization (MO-PPO) algorithm, is proposed. In this algorithm, a loss function-based update strategy is the core point, which is able to calculate weights for both loss functions of coverage and capacity by a min-norm solver at each update. The numerical results demonstrate that the investigated update strategy outperforms the fixed weight-based MO algorithms.
\end{abstract}

\vspace{-0.2cm}
\section{Introduction}
\vspace{-0.1cm}
For supporting increasing heterogeneous quality-of-service requirements of future wireless networks, an emerging communication paradigm, i.e., simultaneously transmitting and reflecting RISs (STAR-RISs) \cite{IEEEhowto:YLiu2} becomes appealing. In contrast to conventional RISs \cite{IEEEhowto:YLiu} , STAR-RISs are able to transmit and reflect the incident signal simultaneously, to achieve full space coverage. Therefore, it is an ultra-interesting question how STAR-RISs perform in terms of coverage and capacity. Note that coverage and capacity optimization (CCO) is one of the typical operational tasks mentioned by the 3rd Generation Partnership Project (3GPP) \cite{3GPP}. Since the coverage and capacity have several conflicting relationships, simultaneously optimizing them is important. For example, high transmit power contributes to large coverage but high inter-cell interference reduces the capacity performance. To this end, multi-objective machine learning (MOML) \cite{IEEEhowto:EBalevi} can be a potential solution.
\par
Conventional performance optimization for STAR-RISs assisted networks focuses on a single objective: capacity or coverage. For capacity performance, there are some primary works. In \cite{Aldababsa2021}, a partitioning algorithm was proposed to determine the proper number of transmitting/reflecting elements that need to be assigned to each user, and maximize the system sum-rate. In STAR-RISs assisted non-orthogonal multiple access systems, the authors of \cite{Zuo2021} proposed a suboptimal two-layer iterative algorithm to maximize the achievable sum rate. For coverage performance, only one recent work has discussed its optimization problem. The STAR-RISs assisted two-user communication networks were studied in \cite{Wu12021}, where the one-dimensional search-based algorithms were proposed to obtain the optimal coverage range. There are mainly three CCO methods based on MOML: 1) Keep one objective in the objective function and move the rest objectives to constraints, while the obtained results are usually sub-optimal. 2) Assign a fixed weight to each objective. This method achieves the optimal results in a single scenario, while it cannot be used in other weight combinations, i.e., other network operation designs. 3) Obtain a set of optimal solutions according to Pareto-based multi-objective optimization algorithms, where one of these solutions can be selected to meet any specific optimization requirement. For the first method, an reinforcement learning (RL)-based solution for coverage and capacity optimization using base station antenna electrical tilt in mobile networks was proposed in \cite{Dandanov2017}. For the second method, in \cite{Skocaj2022}, a minimization of drive tests-driven deep RL algorithm was investigated to optimize coverage and capacity with fixed weights. For the third method, the authors in \cite{Dreifuerst2021} developed and compared two RL-based approaches for maximizing coverage and capacity.
\par
As can be seen from related works, the CCO of STAR-RISs assisted wireless networks is still at its early stage. Modeling the STAR-RISs assisted networks for coverage and capacity and exploring the RL-based solutions of simultaneously optimizing the coverage and capacity are still challenges. To solve these challenges and fully reap the advantages of STAR-RISs, in this paper, we propose a new ML-based on proximal policy optimization (PPO), named multi-objective PPO (MO-PPO), to provide the maximal coverage and capacity for STAR-RISs assisted networks. The main contributions of this paper can be summarized as follows: 1) We propose a new model for a narrow-band downlink STAR-RISs assisted network and formulates the CCO problem of STAR-RIS assisted networks by jointly optimizing the transmit power, the phase shift matrix to solve CCO problem. 2) We adopt a loss function-based update strategy for the MO-PPO algorithm, which is capable of simultaneously obtaining maximum coverage and capacity. 3) We demonstrated that the loss function-based update strategy MO-PPO algorithms are achieving higher benefits than the benchmarks.

\vspace{-0.1cm}
\section{System Model and Problem Formulation}
\vspace{-0.1cm}

\begin{figure*}[htbp]
  \setlength{\belowcaptionskip}{-0.7cm}
  \centering
  \includegraphics[scale = 0.25]{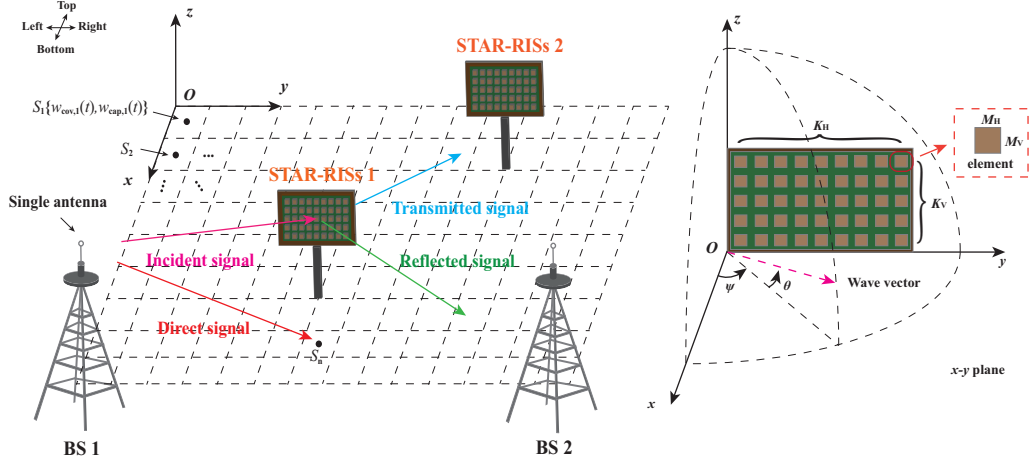}
  \caption{Illustration of the considered STAR-RISs assisted networks.}  
  \label{system_model}
\end{figure*}

As shown in Fig.~\ref{system_model}, we consider a narrow-band downlink STAR-RISs assisted network consisting of two single-antenna BSs and $N_s$ STAR-RISs of the same size equipped with $K = K_HK_V$ reconfigurable elements, where $K_H$ and $K_V$ denote the number of elements per row and column, respectively. The serving range is defined as a square region with the length of the side $R_s$. The BSs are located at the bottom left and bottom right corners with the same height $h_b$, while STAR-RISs with the height $h_{n_s}$ are deployed at designated locations in the square region. Assuming a three-dimensional (3D) Cartesian coordinate system, where the origin is set at the top-left corner. The locations of two BSs and $n_s$-th STAR-RISs are denoted by $\mathrm{B}_1 = (R_s, 0, h_b)$, $\mathrm{B}_2 = (R_s, R_s, h_b)$, and $\mathrm{A}_{n_s} = (x_{n_s},y_{n_s},h_{n_s})$, respectively. Note that $h_{n_s}$ is the height of each STAR-RISs module (including its stand and STAR-RISs), which is far lower than the height of BSs. Therefore, there is a direct link between each BS and any given sampling point. To characterize the coverage and capacity, the region is discretized into numerous square grids with the length of the side $R_g$, while the center point of each grid acts as the sample point. Accordingly, the total number of grids is $N = \lceil R_s/R_g \rceil^2$, where the set of sample points can be denoted as $\mathbf{s} = \{s_1,s_2,...,s_{N}\}$. In practical networks, in order to characterize the importance of each grid at each timestep $t$, two time-related weights, $w_{\mathrm{cov}, i}(t)$ and $w_{\mathrm{cap}, i}(t)$, are assigned for coverage and capacity of each sample points $s_i$ ($i\in [1,N]$), respectively. Moreover, the weights have been unified, i.e., $\sum_{i=1}^{N}w_{\mathrm{cov}, s_i}(t) = 1$ and $\sum_{i=1}^{N}w_{\mathrm{cap}, s_i}(t) = 1$. In this system model, we study to achieve long-term communication with a time period $\mathcal{T}$. For each sample point at any timestep, the weighted assignments $w_{\mathrm{cov}, s_i}(t)$ and $w_{\mathrm{cap}, s_i}(t)$ are influenced by the previous network performance and resource allocation strategy. Therefore, the considered problem can be regarded as a Markov Decision Process (MDP). 

\vspace{-0.2cm}
\subsection{Spatially Correlated Channel Model}
In this section, the fading channels from BSs to STAR-RISs, from STAR-RISs to sample points, and from BSs to sample points are introduced, as well as their spatial channel correlations. Denote $\mathbf{\Phi}_{\delta, n_s}$ as the coefficients of $n_s$-th STAR-RISs with mode $\delta$, where $\delta \in \{\mathrm{Re},\mathrm{Tr}\}$ represents the reflection and transmission modes. Due to the high path loss, this work assume that signals are only reflected and transmitted by the STAR-RISs once. We consider the non-ideal STAR-RISs with same constant amplitude and continuous phase shifters in each mode, where the phase shifters can be expressed as: $\phi_{\delta,n_s,k} \in  [0, 2\pi), \forall k \in \{1,2,\cdots,K\}$. The coefficients of $n_s$-th STAR-RISs are denoted as $\mathbf{\Phi}_{\delta, n_s} = \mathrm{diag}\left(\sqrt{\beta_{\delta,n_s}}e^{j\phi_{\delta,n_s,1}}, ..., \sqrt{\beta_{\delta,n_s}}e^{j\phi_{\delta,n_s,K}}\right), \forall k \in \{1,2,\cdots,K\}$, where $\sqrt{\beta_{\delta,n_s}} \in (0, 1],\hspace{0.5em} \beta_{\mathrm{Re},n_s} + \beta_{\mathrm{Tr},n_s} = 1$. As shown in Fig.~\ref{system_model}, a spherical coordinate system is defined with azimuth angel $\psi$ and elevation angel $\theta$ based on the 3D space. Denote the area of each element as $M = M_HM_V$, where $M_H$ and $M_V$ are the horizontal width and vertical height, respectively. Thus, the total area of $K$ elements can be expressed as $M_a = KM$. For the $k$-th element, its location can be expressed as \cite{r}:
\vspace{-0.1cm}
\begin{align}\label{4}
  l_{k} = [0, x(k)M_H, y(k)M_V]^T,
\end{align}
\par
\vspace{-0.1cm}
\noindent
where $x(k)$ = mod($k-1, K_H$) and $y(k)$ = $\lfloor (k-1)/K_H \rfloor$ are the indices of $k$-th element. Mod($\cdot,\cdot$) and $\lfloor \cdot \rfloor$ denote the modulus operation and truncates the argument. Assume a plane wave with wavelength $\lambda$ is impinging on the STAR-RISs, the array response vector is then given by:
\vspace{-0.1cm}
\begin{align}\label{5}
  \mathbf{a}(\psi, \theta) = [e^{j\mathrm{b}(\psi, \theta)^{T}l_{1}},e^{j\mathrm{b}(\psi, \theta)^{T}l_{2}},\cdots,e^{j\mathrm{b}(\psi, \theta)^{T}l_{K}}]^T,
\end{align}
\par
\vspace{-0.2cm}
\noindent
where $\mathbf{b}(\psi, \theta) \in \mathbb{R}^{3 \times 1}$ is the wave vector, which can be expressed as follows:
\vspace{-0.1cm}
\begin{align}\label{6}
  \mathbf{b}(\psi, \theta) = \frac{2\pi}{\lambda}[\cos(\theta)\cos(\psi), \cos(\theta)\sin(\psi), \sin(\theta)]^T.
\end{align}
\par
\vspace{-0.2cm}
Assume that these channels are independently distributed and corresponding channel state information are perfect. Denote $\mathbf{h}_{a,n_s}$, $\mathbf{h}_{\delta,n_s,s_i}$, and $\mathbf{h}_{a,s_i}$ as the channel from $a$-th BS to $n_s$-th STAR-RISs, from $n_s$-th STAR-RISs to $s_i$-th sample point with mode $\delta$, and from $a$-th BS to $s_i$-th sample point, respectively. Here, the channels $\mathbf{h}_{a,n_s}$, $\mathbf{h}_{\delta,n_s,s_i}$, and $\mathbf{h}_{a,s_i}$ can be modeled as Rician fading model, which are expressed as:
\vspace{-0.1cm}
\begin{align}\label{70}
  \mathbf{h}_{a,n_s} = \sqrt{L_{a\mathrm{R}}} \Big( \sqrt{\frac{\alpha_{a\mathrm{R}}}{1+\alpha_{a\mathrm{R}}}}\mathbf{h}_{a,n_s}^{\mathrm{LOS}} + \sqrt{\frac{1}{1+\alpha_{a\mathrm{R}}}}\mathbf{h}_{a,n_s}^{\mathrm{NLOS}} \Big),
\end{align}
\vspace{-0.6cm}
\begin{align}\label{71}
  \mathbf{h}_{\delta,n_s,s_i} = \sqrt{L_{\mathrm{RP}}} \Big( \sqrt{\frac{\alpha_{\mathrm{RP}}}{1+\alpha_{\mathrm{RP}}}}\mathbf{h}_{n_s,s_i}^{\mathrm{LOS}} + \sqrt{\frac{1}{1+\alpha_{\mathrm{RP}}}}\mathbf{h}_{n_s,s_i}^{\mathrm{NLOS}} \Big),
\end{align}
\vspace{-0.6cm}
\begin{align}\label{72}
  \mathbf{h}_{a,s_i} = \sqrt{L_{a\mathrm{P}}} \Big( \sqrt{\frac{\alpha_{a\mathrm{P}}}{1+\alpha_{a\mathrm{P}}}}\mathbf{h}_{a,s_i}^{\mathrm{LOS}} + \sqrt{\frac{1}{1+\alpha_{a\mathrm{P}}}}\mathbf{h}_{a,s_i}^{\mathrm{NLOS}} \Big),
\end{align}
\par
\vspace{-0.2cm}
\noindent
where $L_{(\mathrm{u})}$ and $\alpha_{(\mathrm{u})}, \mathrm{u} \in \{a\mathrm{R},\mathrm{RP},a\mathrm{P}\}$ denote the corresponding path loss and Rician factor, respectively. $h_{a,s_i}^{\mathrm{LOS}}$ $\sim$ $\mathcal{CN}(0, 1)$ denotes the Rayleigh fading-modeled deterministic line-of-sight (LoS) component of the channel from $a$-th BS to $s_i$-th sample point, while $\mathbf{h}_{a,n_s}^{\mathrm{LOS}} = \mathbf{b}(\psi^{a\mathrm{R}}, \theta^{a\mathrm{R}}) = \mathbf{b}\{\mathrm{arcsin}[ (h_{b}-h_{n_s}) /d_{a, n_s} ], \mathrm{arccos}[(R_s-x_{n_s})/\overline{d}_{a,n_s}]\}$ and $\mathbf{h}_{n_s,s_i}^{\mathrm{LOS}} = \mathbf{b}(\psi^{\mathrm{RP}}, \theta^{\mathrm{RP}}) = \mathbf{b}\{\mathrm{arcsin}(h_{n_s} /d_{n_s,s_i}), \mathrm{arccos}[(x_{n_s}-x_{s_i})/\overline{d}_{n_s,s_i}]\}$ are the deterministic LoS components for the channels from $a$-th BS to $n_s$-th STAR-RISs, and from $n_s$-th STAR-RISs to $s_i$-th sample point, respectively. Among them, $d_{a,n_s}$ and $d_{n_s,s_i}$ denote 3D distance between $a$-th BS and $n_s$-th STAR-RISs, and 3D distance between $n_s$-th STAR-RISs and $s_i$-th sample point, while $\overline{d}_{a,n_s}$ and $\overline{d}_{n_s,s_i}$ denote 2D distance between $a$-th BS and $n_s$-th STAR-RISs, and 2D distance between $n_s$-th STAR-RISs and $s_i$-th sample point. $x_{n_s}$, $x_{s_i}$ indicate the $n_s$-th STAR-RISs, and $s_i$-th sample point, respectively. $\mathbf{h}_{a,n_s}^{\mathrm{NLOS}} \sim \mathcal{CN}\big(0, \mathbb{E}\big[\mathbf{h}_{a,n_s}^{\mathrm{NLOS}}(\mathbf{h}_{a,n_s}^{\mathrm{NLOS}})^{H}\big]\big)$, $\mathbf{h}_{n_s,s_i}^{\mathrm{NLOS}} \sim \mathcal{CN}\big(0, \mathbb{E}\big[\mathbf{h}_{n_s,s_i}^{\mathrm{NLOS}}(\mathbf{h}_{n_s,s_i}^{\mathrm{NLOS}})^{H}\big]\big)$, and $\mathbf{h}_{a,s_i}^{\mathrm{NLOS}} \sim \mathcal{CN}(0, 1)$ are the non-line-of-sight (NLoS) components modeled as Rayleigh fading. Furthermore, for path loss $L_{(\mathrm{u})}$, it can be modeled as $L_{u} = Cd_\mathrm{v}^{-\gamma_\mathrm{v}}, \mathrm{v} \in \{\{a,n_s\},\{n_s,s_i\},\{a,s_i\}\}$, where $C$ denotes the path loss at the reference distance of 1 meter and $\gamma_\mathrm{v}$ represents the path loss factor.

\vspace{-0.2cm}
\subsection{Signal Model}
\vspace{-0.1cm}
Since the size of the STAR-RISs module affect the direct link. The received signal $y_{a,n_s,s_i} \in \mathbb{C}$ from the $a$-th BS to the $s_i$ sample point via $n_s$-th STAR-RISs can be written as:
\vspace{-0.1cm}
\begin{align}\label{signal model}
  y_{a,n_s,s_i} = 
      \left(\mathbf{h}_{\delta,n_s,s_i}^\mathrm{H} \mathbf{\Phi}_{\delta, n_s} \mathbf{h}_{a,n_s} + \mathbf{h}_{a,s_i}\right)x + n_{a,n_s,s_i},
\end{align}
\par
\vspace{-0.2cm}
\noindent
where the total transmit power $P_t = |x|^2$ and $n_{a,n_s,s_i} \sim \mathcal{CN}(0, \sigma^2)$ is the additive white Gaussian noise variance. Based on the received signal power, the reference signal receiving power (RSRP) can be defined as the maximal useful signal power from all possible sources. The RSRP at the sample point $s_i$ is given by:
\vspace{-0.1cm}
\begin{align}\label{8}
\mathrm{RSRP}_{s_i} = \max\limits_{a \in \{1, 2\}, n_s \in \{1, 2, \cdots, N_s\}} |y_{a,n_s,s_i}|^2.
\end{align}
\par
\vspace{-0.2cm}
The achievable signal-to-interference-plus-noise ratio (SINR) of $s_i$ sample point is calculated as follows:
\vspace{-0.1cm}
\begin{align}\label{9}
  \mathrm{SINR}_{a,n_s,s_i}  = \frac{|y_{a,n_s,s_i} - n_{a,n_s,s_i}|^2}{ \sum_{n_s^{'}=1,n_s^{'}\neq n_s}^{|\mathbf{S}|} |y_{a^{'},n_s^{'},s_i}-n_{a^{'},n_s^{'},s_i}|^2+{n_{a,n_s,s_i}^2}},
\end{align}
\par
\vspace{-0.2cm}
\noindent
where $a=1$, $a^{'}=2$; and $a=2$, $a^{'}=1$, otherwise. Assume the minimal RSRP for all sample points is $\mathrm{R}_{th}$, the weighted coverage ratio at time $t$ can be written as
\vspace{-0.1cm}
\begin{align}\label{10}
  \mathrm{Coverage}(t) = \frac{|\mathbf{w}_{\mathrm{cov}, \check{\mathbf{s}}(t)} \cdot \check{\mathbf{s}}(t)|}{N},
\end{align}
\par
\vspace{-0.2cm}
\noindent
where $\check{\mathbf{s}}(t) = \{\check{s}_1(t),\check{s}_2(t),\cdots,\check{s}_{\tilde{N}}(t)\}$ is the set of the sample points at time $t$ that satisfying the condition $\mathrm{RSRP}_{\check{s}_{\tilde{n}}(t)} \geq \mathrm{R}_{th}, \check{s}_{\tilde{n}}(t) \in \check{\mathbf{s}}(t)$. $\mathbf{w}_{\mathrm{cov}, \check{\mathbf{s}}}(t) = \{w_{\mathrm{cov}, \check{s}_1}(t), w_{\mathrm{cov}, \check{s}_2}(t), \cdots, w_{\mathrm{cov}, \check{s}_{\tilde{N}}}(t)\}$ is the normalized corresponding coverage weights for the sample points $\check{\mathbf{s}}(t)$. For the network capacity, it is mainly determined by SINR, so at the time $t$, the weighted capacity can be represented by
\vspace{-0.1cm}
\begin{align}\label{11}
  \mathrm{Capacity}(t)  = \sum_{s_i=1}^{N_s} w_{\mathrm{cap}, s_i}(t) \cdot \mathrm{B} \log_2\left(1+\mathrm{SINR}_{a^*,k^*,s_i}(t) \right),
\end{align}
\par
\vspace{-0.2cm}
\noindent
where $\mathrm{B}$ is the system bandwidth and $a^*, n_s^*= \arg\max_{a \in \{1, 2\}, n_s \in \{1, 2, \cdots, N_s\}} |y_{a,n_s,s_i}|^2$.

\subsection{Problem Formulation}
We focus on maximizing the long-term coverage and capacity by optimizing the transmit power, the reflection phase shift matrix, the transmission phase shift matrix, and time $\mathcal{T}$. The formulated problem can be expressed as follows:
\vspace{-0.3cm}
\begin{align}
  &\underset{P_t, \mathbf{\Phi}_{\mathrm{Re}, n_s},  \mathbf{\Phi}_{\mathrm{Tr}, n_s}, \mathcal{T}}{\max} \hspace*{1em} \sum_{t=1} ^ \mathcal{T} \big[\mathrm{Coverage}(t), \mathrm{Capacity}(t)\big] \label{12}\\
  &\mathrm{s.\ t.} \hspace*{1em} 0<P_t\le P_{t,\mathrm{max}},\tag{\ref{12}{a}} \label{12a}\\
  & \hspace*{2.75em} 0< \mathrm{tr}(\mathbf{\Phi}^H_{\delta,n_s}\mathbf{\Phi}_{\delta,n_s})< 1, \tag{\ref{12}{b}} \label{12b}\\
  & \hspace*{2.75em} 0<\mathrm{tr}(\mathbf{\Phi}^H_{\mathrm{Re},n_s}\mathbf{\Phi}_{\mathrm{Re},n_s}) + \mathrm{tr}(\mathbf{\Phi}^H_{\mathrm{Tr},n_s}\mathbf{\Phi}_{\mathrm{Tr},n_s})\le 1, \tag{\ref{12}{c}} \label{12c}
\end{align}

\par
\vspace{-0.3cm}
\noindent
where $P_{\mathrm{max}}$ and $\mathbf{C} \subset \mathbb{R}^2$ denote the permitted maximum transmit power and the considered serving area, respectively. Constraint \eqref{12a} limits the range of the transmit power. According to the energy conservation principle, constraints \eqref{12b} and \eqref{12c} show that both the energy of different modes and the sum energy of the reflected and transmitted signals is less than one. However, the main difficulty in solving the problem \eqref{12} owing to the following reasons. Firstly, the NLoS components for STAR-RISs assisted links are hard to be determined before the STAR-RISs deployment, where the locations of STAR-RISs are infinite and rely on the no concave distribution of coverage and capacity of each sample point. Secondly, the distribution weights $w_{\mathrm{cov}, s_i}(t)$ and $w_{\mathrm{cap}, s_i}(t)$ at time $t$ for calculating the coverage and capacity is not a continuous function. Thirdly, with respect to the continuous-time $t$, it's difficult to handle infinite variables optimization, since any adjacent time is submitted to the Markov chain. Thus, conventional non-convex optimization methods are not suitable for solving these difficulties. In the next section, the Pareto optimal-based MO-PPO algorithm is invoked to solve this problem.

\section{Pareto optimal-based MO-PPO Algorithm}
In this section, we firstly elaborate on the MDP in the MO-PPO algorithm. Then, the update strategy of the Pareto optimal (PO)-based MO-PPO algorithm is proposed to verify the optimal policy applicable to the system model.

\subsection{MO-PPO Framework}
In MO-PPO algorithm, MDP is represented by a tuple $\langle \mathbf{S}, \mathbf{A}, \mathbf{p}, \mathbf{R}\rangle$ with state space $\mathbf{S}$, action space $\mathbf{A}$, and transition probability matrix $\mathbf{p}$. Define a controller as an agent, which can control both two BSs, to develop the policy from the BSs to sample points via STAR-RISs, and phase shifters. At each timestep $t$, the controller can observe the state $\mathbf{S}_t$ from state space $\mathbf{S}$, and carries out an action $\mathbf{A}_t$ from action space $\mathbf{A}$. The received reward is to make the transition to the next state $\mathbf{S}_{t+1}$. In this system model, the locations of STAR-RISs are randomly chosen. Note that, the locations of STAR-RISs are not overlapped. Therefore, the distance between any BS and $s_i$-th sample point is fixed, while the coverage and capacity are mainly determined by the distance between STAR-RISs and $s_i$-th point and the corresponding phase shift of the STAR-RISs, according to the \eqref{10} and \eqref{11}. Thus, the state $\mathbf{S}_{t}$ can be defined symbolically as follows:
\vspace{-0.1cm}
\begin{align}\label{state space}
  \mathbf{S}_t = 
  \begin{bmatrix} 
    \beta_{\mathrm{Re},n_s}(t), \beta_{\mathrm{Tr},n_s}(t), \mathbf{\Phi}_{\mathrm{Re},n_s}(t), \mathbf{\Phi}_{\mathrm{Tr},n_s}(t)
  \end{bmatrix}.
\end{align}
\par
\vspace{-0.2cm}
For the action $\mathbf{A}_t$, the $\beta_{\mathrm{Tr},n_s}$ of STAR-RISs is discreted with small step $z$ as numerous values between $(0, 1)$, while the $\beta_{\mathrm{Re},n_s}$ is determined by $(1-\beta_{\mathrm{Tr},n_s})$. The phase shifters follows the continuous phase definition $[0, 2\pi)$ of STAR-RISs.
\vspace{-0.1cm}
\begin{align}\label{action space}
  \mathbf{A}_t = 
  \begin{bmatrix} 
    \Delta \beta_{\mathrm{Re},n_s}, \Delta \beta_{\mathrm{Tr},n_s}, \Delta \phi_{\mathrm{Re},n_s}, \Delta \phi_{\mathrm{Tr},n_s}
  \end{bmatrix}.
\end{align}
\par
\vspace{-0.2cm}
\noindent
where $\Delta \beta_{\mathrm{Re},n_s} \in \{z,2z,\cdots,1-z\}$, $\Delta \beta_{\mathrm{Tr},n_s} \in \{1-z,1-2z,\cdots,z\}$ and $\Delta \phi_{\delta,n_s} = \{ \phi_{\delta,n_s,1},\phi_{\delta,n_s,2},\cdots,\phi_{\delta,n_s,K}\}$ denote the possible values for the tranmission amplitude, reflection amplitude, and possible phases for $n_s$-th STAR-RISs with mode $\delta$, respectively. For $k$-th element, the phase is randomly selected from $[0,2\pi)$. To obtain the maximum transmission coverage and capacity that BSs can achieve in a time period $\mathcal{T}$, the reward is denoted as the difference of coverage and capacity in adjuscent timesteps, which can be expressed as:
\vspace{-0.1cm}
\begin{align}\label{Multi-objective reward}
  \mathbf{R}_t(\mathbf{S}_t, \mathbf{A}_t) = 
  \begin{bmatrix}
    \Delta \mathrm{Cov}_{t \rightarrow t+1}, \Delta \mathrm{Cap}_{t \rightarrow t+1}
  \end{bmatrix}.
\end{align}
\par
\vspace{-0.2cm}
For the loss function in the PPO algorithm, there are two Approaches: The clipped surrogate objective and the Adaptive KL penalty coefficient, which can be used as an evaluation of the loss function.

\vspace{-0.3cm}
\subsection{Loss Function-based Update Strategy}
In this work, we consider an update strategy for the Pareto optimal-based MO-PPO algorithm, i.e, the loss function-based update strategy, where the multi-task learning (MTL) method is employed. Different from the conventional update strategy, here are multiple gradient policies that need to be updated simultaneously. In the MTL-based MO-PPO problem, the empirical risk minimization formulation is generally followed:
\vspace{-0.1cm}
\begin{align}\label{empirical risk minimization}
  \min_{\pmb{\overline{\theta}}} \sum_{m=1}^{M} \varphi^m \hat{\mathcal{L}}^m(\pmb{\overline{\theta}}),
\end{align}
\par
\vspace{-0.2cm}
\noindent
where $\varphi^m$ and $\hat{\mathcal{L}}^m(\pmb{\overline{\theta}})$ denote the weights for $m$-th task and the empirical loss of $m$-th task. Consider two sets of solutions $\pmb{\overline{\theta}}_1$ and $\pmb{\overline{\theta}}_2$, if $\hat{\mathcal{L}}^{1}(\pmb{\overline{\theta}}_1) > \hat{\mathcal{L}}^{1}(\pmb{\overline{\theta}}_2)$ and $\hat{\mathcal{L}}^{2}(\pmb{\overline{\theta}}_1) < \hat{\mathcal{L}}^{2}(\pmb{\overline{\theta}}_2)$, it can be obtained that the two tasks are mutually non-dominated, and therefore belong to the Pareto front. In this case, MTL problem can be formulated as MO optimization to explore the optimal results for conflicting objectives, where the vector-valued loss $\pmb{\mathcal{L}}$ are employed as follows:
\vspace{-0.1cm}
\begin{align}\label{vector loss}
  \min_{\overline{\pmb{\overline{\theta}}}} \pmb{\mathcal{L}}(\overline{\pmb{\overline{\theta}}}) = \min_{\pmb{\overline{\theta}}} [\hat{\mathcal{L}}^{1}(\pmb{\overline{\theta}}), \hat{\mathcal{L}}^{2}(\pmb{\overline{\theta}}), \cdots, \hat{\mathcal{L}}^{M}(\pmb{\overline{\theta}})]^T.
\end{align}
\par
\vspace{-0.2cm}
Hence, the optimization of equation \eqref{vector loss} is to find PO solutions. Define $\mathcal{F}=\{\pmb{\mathcal{L}}(\pmb{\overline{\theta}})\}, \pmb{\overline{\theta}} \in \pmb{\overline{\Theta}}$ as the Pareto front, where $\pmb{\overline{\theta}}$ and $\pmb{\overline{\Theta}}$ denote the any one set of optimal parameters and all possible sets of optimal parameters. 

\subsubsection{Multiple Gradient Descent Algorithms (MGDA)}
Multiple gradient descent algorithms (MGDA) \cite{MGDA} is a proper method to converge to the Pareto stationary solution problem. According to the Karush-Kuhn-Tucker (KKT) conditions, there exists $\nu_1,\nu_2,\cdots,\nu_M$ such that:
\begin{itemize}
  \item  $\nu_1,\nu_2,\cdots,\nu_M \geq 0$.
  \item  $\sum_{m=1}^{M}\nu_m = 1$ and $\sum_{m=1}^{M}\nu_m \nabla_{\pmb{\overline{\theta}}} \hat{\mathcal{L}}^m(\pmb{\overline{\theta}}) = 0$.
\end{itemize}
\par
Before handling the MGDA algorithms, the objectives may have values of the different scales, while MGDA is sensitive to the different ranges. Thus, the following gradient normalization method is invoked to alleviate the value range:
\vspace{-0.1cm}
\begin{align}\label{normalization}
  \nabla_{\pmb{\overline{\theta}}}\pmb{\mathcal{L}}(\pmb{\overline{\theta}}) = \frac{\nabla_{\pmb{\overline{\theta}}}\pmb{\mathcal{L}}(\pmb{\overline{\theta}})}{\pmb{\mathcal{L}}(\hat{\pmb{\overline{\theta}}})},
\end{align}
\par
\vspace{-0.2cm}
\noindent
where $\hat{\pmb{\overline{\theta}}}$ is the initial parameters of the model. Consequently, the range of loss function has been limited to $[0, 1]$. 

\begin{definition}\label{definition 1}
  A solution $\pmb{\overline{\theta}}_1$ dominates a solution $\pmb{\overline{\theta}}_2$ if for all objectives satisfying $\hat{\mathcal{L}}^{m}(\pmb{\overline{\theta}}_1) \leq \hat{\mathcal{L}}^{m}(\pmb{\overline{\theta}}_2)$, while exists at least one objective satisfying $\hat{\mathcal{L}}^{n}(\pmb{\overline{\theta}}_1) < \hat{\mathcal{L}}^{n}(\pmb{\overline{\theta}}_2)$, $\forall m,n \in \{1,2,\cdots,M\}$.
\end{definition}

\begin{definition}\label{definition 2}
  A solution $\pmb{\overline{\theta}}_1$ is PO solution while there is no any other solution $\pmb{\overline{\theta}}_2$ dominates $\pmb{\overline{\theta}}_1$.
\end{definition}

\begin{definition}\label{definition 3}
  All non-dominated solutions $\hat{\pmb{\overline{\theta}}}$ are Pareto set.
\end{definition}

The solution that satisfies the conditions above is defined as a Pareto stationary solution, while the Pareto optimal solution is Pareto stationary solution. Since it has two objectives in problem \eqref{12}, the optimization problem can be defined as follows:
\vspace{-0.1cm}
\begin{align}\label{QCOP2}
  \min_{\nu \in [0,1]} ||\nu \nabla_{\pmb{\overline{\theta}}} \hat{\mathcal{L}}^1(\pmb{\overline{\theta}}) + (1-\nu) \nabla_{\pmb{\overline{\theta}}} \hat{\mathcal{L}}^2(\pmb{\overline{\theta}})||^2_2,
\end{align}
\par
\vspace{-0.2cm}
\noindent
where $||\cdot||^2_2$ and $\nabla_{[\cdot]}$ denote the L2 norm and gradient descent (GD) operator. Define $\nabla_{\overline{\pmb{\overline{\theta}}}} \mathcal{L}(\overline{\pmb{\overline{\theta}}}) = \sum_{m=1}^{M}\nu_m \nabla_{\pmb{\overline{\theta}}} \hat{\mathcal{L}}^m(\pmb{\overline{\theta}})$, we have that: if $\nabla_{\overline{\pmb{\overline{\theta}}}} \mathcal{L}(\overline{\pmb{\overline{\theta}}}) = 0$, the solution is Pareto stationary; otherwise, it isn't Pareto stationary and $\nabla_{\overline{\pmb{\overline{\theta}}}} \mathcal{L}(\overline{\pmb{\overline{\theta}}})$ is the general GD vector. The optimization problem defined in \eqref{QCOP2} is equivalent to finding a minimum-norm point in the convex hull, which is a convex quadratic problem with linear constraints. Thus, an analytical solution to equation \eqref{QCOP2} can be expressed as:
\vspace{-0.3cm}
\begin{align}\label{QCOP2-solution}
  \nu = \bigg\{ \frac{[\nabla_{\pmb{\overline{\theta}}} \hat{\mathcal{L}}^2(\pmb{\overline{\theta}}) - \nabla_{\pmb{\overline{\theta}}} \hat{\mathcal{L}}^1(\pmb{\overline{\theta}})]^T\nabla_{\pmb{\overline{\theta}}} \hat{\mathcal{L}}^2(\pmb{\overline{\theta}})}{||\nabla_{\pmb{\overline{\theta}}} \hat{\mathcal{L}}^1(\pmb{\overline{\theta}}) - \nabla_{\pmb{\overline{\theta}}} \hat{\mathcal{L}}^2(\pmb{\overline{\theta}})||^2_2} \bigg\}_{[0,1]},
\end{align}
\par
\vspace{-0.2cm}
\noindent
where $\{\}_{[0,1]}$ represents clipping $\nu$ to $[0,1]$. Alternate optimization of GD vector and $\nu$ produces different $\nu$, which covers all Pareto optimal solutions under constraints to form Pareto frontiers. According to the system model, it's suitable to select one PO solution as the optimal result. Therefore, we select the worse objective value optimized by the Pareto optimal solution between two objectives for comparison, and the smaller one is the final desired optimal solution.


\subsubsection{Loss Function}
Our goal is to train one policy containing two sub-policies, where each objective has a specific loss function and shares all parameters. Thus, combing with the PPO algorithm, the loss function for the MO-PPO algorithm based on the No clipping or penalty method, clipped method, and KL Penalty method can be expressed as \eqref{loss function1}, \eqref{loss function2}, and \eqref{loss function3}, where $\hat{\mathbf{A}}_t$ is an advantage estimator, it can be expressed as \eqref{advantage}. The pseudo code of the algorithm is shown in \textbf{Algorithm~\ref{loss-MOPPO}}. 
\begin{figure*}[hbp]
\normalsize 
\begin{align}
  &\pmb{\mathcal{L}}^{\mathrm{NCP}}(\pmb{\overline{\theta}}) = \min_{\nu \in [0, 1]} \Bigg|\Bigg| \nu \mathbb{E}_{t}^{1}\Big\{\mathrm{min}\Big[\mathrm{\frac{\pi_{\overline{\theta}^{*}}(\mathbf{S}_t, \mathbf{A}_t)}{\pi_{\overline{\theta}}(\mathbf{S}_t, \mathbf{A}_t)}\hat{\mathbf{A}}_{t}^{\pi^{*}}}\Big]\Big\} + (1-\nu) \mathbb{E}_{t}^{2}\Big\{\mathrm{min}\Big[\mathrm{\frac{\pi_{\overline{\theta}^{*}}(\mathbf{S}_t, \mathbf{A}_t)}{\pi_{\overline{\theta}}(\mathbf{S}_t, \mathbf{A}_t)}\hat{\mathbf{A}}_{t}^{\pi^{*}}}\Big]\Big\}\Bigg|\Bigg|^{2}_{2}, \label{loss function1}\\ 
  &\pmb{\mathcal{L}}^{\mathrm{CLIP}}(\pmb{\overline{\theta}}) = \min_{\nu \in [0, 1]} \Bigg|\Bigg| \nu \mathbb{E}_{t}^{1}\Big\{\mathrm{min}\Big[\mathrm{\frac{\pi_{\overline{\theta}^{*}}(\mathbf{S}_t, \mathbf{A}_t)}{\pi_{\overline{\theta}}(\mathbf{S}_t, \mathbf{A}_t)}\hat{\mathbf{A}}_{t}^{\pi^{*}},clip\Big(\frac{\pi_{\overline{\theta}^{*}}(S_t, A_t)}{\pi_{\overline{\theta}}(S_t, A_t)}, 1-\epsilon, \epsilon\Big)\hat{\mathbf{A}}_{t}^{\pi^{*}}}\Big]\Big\} \nonumber \\ 
  &\hspace{16em} + (1-\nu) \mathbb{E}_{t}^{2}\Big\{\mathrm{min}\Big[\mathrm{\frac{\pi_{\overline{\theta}^{*}}(\mathbf{S}_t, \mathbf{A}_t)}{\pi_{\overline{\theta}}(\mathbf{S}_t, \mathbf{A}_t)}\hat{\mathbf{A}}_{t}^{\pi^{*}},clip\Big(\frac{\pi_{\overline{\theta}^{*}}(S_t, A_t)}{\pi_{\overline{\theta}}(S_t, A_t)}, 1-\epsilon, \epsilon\Big)\hat{\mathbf{A}}_{t}^{\pi^{*}}}\Big]\Big\}\Bigg|\Bigg|^{2}_{2}, \label{loss function2}\\ 
  &\pmb{\mathcal{L}}^{\mathrm{KL}}(\pmb{\overline{\theta}}) = \min_{\nu \in [0, 1]} \Bigg|\Bigg| \nu \mathbb{E}_{t}^{1}\Big\{\mathrm{min}\Big[\mathrm{\frac{\pi_{\overline{\theta}^{*}}(\mathbf{S}_t, \mathbf{A}_t)}{\pi_{\overline{\theta}}(\mathbf{S}_t, \mathbf{A}_t)}\hat{\mathbf{A}}_{t}^{\pi^{*}}, \tilde{\beta}KL(\pi_{\overline{\theta}^{*}}(\mathbf{S}_t), \pi_{\overline{\theta}}(\mathbf{S}_t))}\Big]\Big\} \nonumber \\
  &\hspace{16em} + (1-\nu) \mathbb{E}_{t}^{2}\Big\{\mathrm{min}\Big[\mathrm{\frac{\pi_{\overline{\theta}^{*}}(\mathbf{S}_t, \mathbf{A}_t)}{\pi_{\overline{\theta}}(\mathbf{S}_t, \mathbf{A}_t)}\hat{\mathbf{A}}_{t}^{\pi^{*}},\tilde{\beta}KL(\pi_{\overline{\theta}^{*}}(\mathbf{S}_t), \pi_{\overline{\theta}}(\mathbf{S}_t))}\Big]\Big\}\Bigg|\Bigg|^{2}_{2}, \label{loss function3}
\end{align}
\hrulefill \vspace*{0pt}
\end{figure*}

\begin{figure*}[hbp]
\normalsize 
\begin{align}\label{advantage}
  \hat{\mathbf{A}}_{t}^{\pi^{*}} &= \sum_{t}^{\overline{T}}\mathbf{Q}_{\pi}(\mathbf{S}_t, \mathbf{A}_t) - V_{\pi}(\mathbf{S}_t) = \mathbf{R}_{t} + \gamma \mathbf{R}_{t+1} + \gamma^2 \mathbf{R}_{t+2} + \cdots + \gamma^{\overline{T}-t+1}\mathbf{R}_{\overline{T}-1} + \gamma^{\overline{T}-t}V_{\pi}(\mathbf{S}_{\overline{T}}) - V_{\pi}(\mathbf{S}_t).
\end{align}
\hrulefill \vspace*{0pt}
\end{figure*}

\vspace{-0.2cm}
\begin{algorithm}[htbp]
  \caption{Pareto optimal-based MO-PPO algorithm, loss function-based update strategy}
  \label{loss-MOPPO}
  \begin{algorithmic}[1]
  \REQUIRE ~~\\
  PPO network structure.\\
  \ENSURE The policy network.\\
  \STATE \textbf{Initialize:} Hyperparameters of PPO network.
  \FOR {iteration = 1, 2, $\cdots$}
    \FOR {objective = 1, 2, $\cdots$}
      \FOR {actor = 1, 2, $\cdots$, N}
        \STATE Run policy $\pi_{\overline{\theta}}$ in environment for $T$ timesteps for each objective.
        \STATE Compute advantage estimates $\hat{A}_{1}, \cdots, \hat{A}_{T}$ for each objective.
      \ENDFOR
    \ENDFOR
    \STATE Calculate loss function $\pmb{\mathcal{L}}$ wrt $\overline{\pmb{\theta}}$, with $\overline{U}$ epochs and minibatch size $M \leq \mathbf{\mathcal{U}}$ update frequency, according to equation \eqref{loss function1}, \eqref{loss function2}, or \eqref{loss function3}.
    \STATE Update $\overline{\pmb{\theta}}$ by min-norm solver.
  \ENDFOR
  \end{algorithmic}
\end{algorithm}


\section{Numerical Results}

In this section, we provide numerical results to evaluate the performance of the proposed MO-PPO algorithm and the explored Pareto-optimal solution. Without loss of generality, a Poisson traffic model is employed to estimate the traffic flows or data sources in the proposed system model.  The hyper-parameters for algorithms training are denoted as the default of the original PPO algorithm \cite{PPO}. Additionally, there are two cases are conceived to help evaluate the proposed update strategies: \textbf{Weights 0.3 and 0.7} and \textbf{Weights 0.6 and 0.4}, which indicates that the weights of coverage and capacity are fixed as 0.3 and 0.7, 0.6 and 0.4, respectively. Then, we discuss the number of STAR-RISs and the number of elements in STAR-RISs on the optimal coverage and capacity.

\begin{figure*}[htbp]
  \setlength{\abovecaptionskip}{-0.1cm}
  \setlength{\belowcaptionskip}{-0.6cm}
  \centering
  \subfigure[The optimized coverage versus different number of STAR-RISs.]
  {
  \begin{minipage}[t]{0.45\textwidth}
  \centering
  \includegraphics[height=2in, width=3.2in]{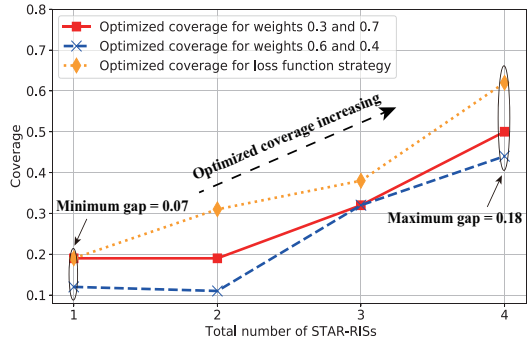}
  \label{optimized_cov_num}
  \end{minipage}
  }\hspace{0.75cm}
  \subfigure[The optimized capacity versus different number of STAR-RISs.]
  {
  \begin{minipage}[t]{0.45\textwidth}
  \centering
  \includegraphics[height=2in, width=3.2in]{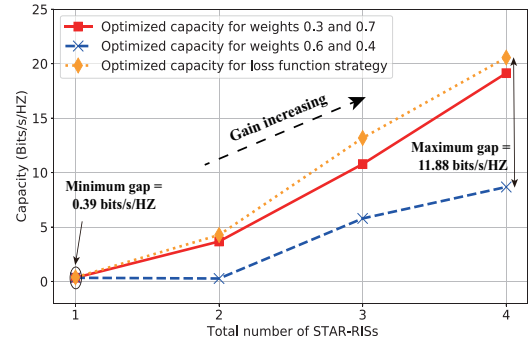}
  \label{optimized_cap_num}
  \end{minipage}
  }
  \caption{The Optimized coverage and capacity for the MO-PPO algorithm with fixed weights, action value-based update strategy, and loss function-based update strategy with different number $N_s$ of STAR-RISs, $K = 8$.}
  \label{optimized_num}
\end{figure*}

\subsubsection{Impact of the Number of STAR-RISs}
Fig.~\ref{optimized_num} depicts the optimized coverage and capacity versus the different numbers of STAR-RISs. In this scenario, the elements $K$ are defined as: $K = 8$. As shown in Fig.~\ref{optimized_cov_num}, the coverage of all cases keeps growing steadily as the number of STAR-RISs increases. When the number of STAR-RISs $N_s$ reaches 4, the coverage of the \textbf{Weights 0.3 and 0.7} and \textbf{Weights 0.6 and 0.4} case can be promoted to over 0.4, and the proposed update strategy can arrive over 0.6. For the capacity depicted in Fig.~\ref{optimized_cap_num}, the gap between the loss function update strategy and \textbf{Weights 0.4 and 0.6} case are enlarged from 0.39 bits/s/HZ to 11.88 bits/s/HZ as the number of STAR-RISs $N_s$ increasing. These are because, with the increase in the number of STAR-RISs, STAR-RISs can help to compensate the received RSRP of some sample points to reach $\mathrm{R}_{th}$, where channels among these sample points and BS are severely attenuated by the distance. Thus, the proposed update strategy outperforms the benchmarks.

\begin{figure*}[htbp]
  \setlength{\abovecaptionskip}{-0.1cm}
  \setlength{\belowcaptionskip}{-0.7cm}
  \centering
  \subfigure[The optimized coverage versus different elements of STAR-RISs.]
  {
  \begin{minipage}[t]{0.45\textwidth}
  \centering
  \includegraphics[height=2in, width=3.2in]{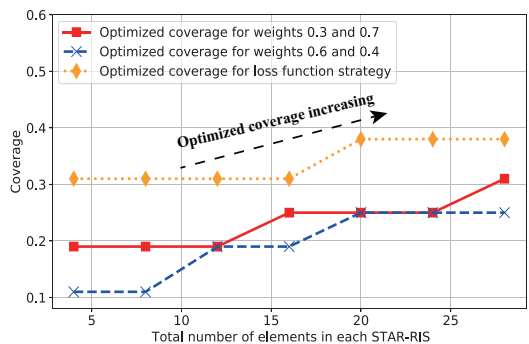}
  \label{optimized_cov_ele}
  \end{minipage}
  }\hspace{0.75cm}
  \subfigure[The optimized capacity versus different elements of STAR-RISs.]
  {
  \begin{minipage}[t]{0.45\textwidth}
  \centering
  \includegraphics[height=2in, width=3.2in]{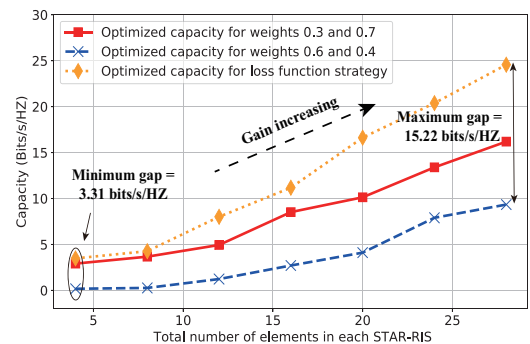}
  \label{optimized_cap_ele}
  \end{minipage}
  }
  \caption{The Optimized coverage and capacity for the MO-PPO algorithm with fixed weights, action value-based update strategy, and loss function-based update strategy with different elements $K$ of each STAR-RISs, $N_s = 2$.}
  \label{optimized_ele}
\end{figure*}

\subsubsection{Impact of the Number of Element in Each STAR-RISs}
Fig.~\ref{optimized_ele} describes the optimized coverage and capacity versus the different number of elements in each STAR-RISs. In this scenario, the number of STAR-RISs $N_s$ is defined as: $N_s = 2$. It can be observed that the coverage shows a slight change in Fig.\ref{optimized_cov_ele}, while the maximum gaps among the optimized capacity of three cases in Fig.\ref{optimized_cap_ele} keep increasing from 3.31 bits/s/HZ to 15.22 bits/s/HZ as the number of elements in each STAR-RISs upgrades. These are because the role of each element is to transmit the BS signal to each sampling point while increasing the number of elements of each STAR-RISs is adding multiple links to reduce loss. Compared with increasing the number of STAR-RISs, increasing the number of elements does not change the channel fast fading characteristics of distant sample points. Therefore, it can be obtained that the proposed update strategy outperforms the benchmarks.

\vspace{-0.2cm}
\section{Conclusion}
\vspace{-0.1cm}
In this paper, we proposed a new model for dynamic CCO in STAR-RISs assisted wireless networks, by optimizing the transmit power and the phase shift matrix. In order to simultaneously optimize the coverage and capacity, a loss function-based update strategy was investigated. The core idea of the proposed strategy was to consider the two-loss function for coverage and capacity, which was dynamically assigned the weights by a min-norm solver at each update. The numerical results proved that when considering the different number of STAR-RISs and the different number of elements in the STAR-RISs, the investigated update strategy outperforms the fixed weight-based MO algorithms.

\end{sloppypar}
\end{document}